\UseRawInputEncoding
\documentclass[
reprint,
superscriptaddress,
amsmath,amssymb,
aps,
prl,
]{revtex4-2}
\usepackage{graphicx}
\usepackage{dcolumn}
\usepackage{mathrsfs}
\usepackage{bm}
\begin{document}
\title{Elastic Softening in Synthetic Diamonds}
\author{Tatsuya Yanagisawa}
 \affiliation{Department of Physics, Hokkaido University, Sapporo 060-0810, Japan}
\author{Ruo Hibino}
 \affiliation{Department of Physics, Hokkaido University, Sapporo 060-0810, Japan}
\author{Hiroyuki Hidaka}
 \affiliation{Department of Physics, Hokkaido University, Sapporo 060-0810, Japan}
\author{Hiroshi Amitsuka}
 \affiliation{Department of Physics, Hokkaido University, Sapporo 060-0810, Japan}
\author{Toshiyuki Tashima}
 \affiliation{Department of Electronic Science and Engineering, Kyoto University, Kyoto 615-8510, Japan}
\author{Mitsuhiro Akatsu}
 \affiliation{Graduate School of Science and Technology, Niigata University, Niigata 950-2181, Japan}
\author{Yuichi Nemoto}
 \affiliation{Graduate School of Science and Technology, Niigata University, Niigata 950-2181, Japan}
\author{Sergei Zherlitsyn}
 \affiliation{Hochfeld-Magnetlabor Dresden (HLD-EMFL) and W\"urzburg-Dresden Cluster of Excellence ct.qmat, Helmholtz-Zentrum Dresden-Rossendorf (HZDR), 01328 Dresden, Germany}
\author{Joachim Wosnitza}
 \affiliation{Hochfeld-Magnetlabor Dresden (HLD-EMFL) and W\"urzburg-Dresden Cluster of Excellence ct.qmat, Helmholtz-Zentrum Dresden-Rossendorf (HZDR), 01328 Dresden, Germany}
 \affiliation{Institut f\"{u}r Festk\"{o}rper- und Materialphysik, TU Dresden, 01062 Dresden, Germany}
\date{\today}

\begin{abstract}
This study reveals a previously unreported phenomenon: elastic softening of synthetic diamonds at temperatures below 1 K. We present ultrasonic measurements on single-crystalline, non-irradiated synthetic diamonds—namely, type-IIa (colorless) and type-Ib (yellow) diamonds grown by high-pressure high-temperature (HPHT) synthesis, as well as type-IIa diamond grown by chemical vapor deposition (CVD). A pronounced, divergent decrease in the elastic stiffness constant $C_{44}$ was observed in all samples down to 20 mK. We attribute this softening to electric quadrupolar degrees of freedom with irreducible representation $T_2$ in diamond. The microscopic origin of this effect, however, remains unresolved. By analogy with similar behavior observed in silicon, we suggest the presence of an as-yet-unidentified, defect-derived quantum ground state with $T_1$ or $T_2$ symmetry at ppb-level concentrations in all three diamonds studied.
\end{abstract}
\maketitle

Diamonds have long attracted interest not only for their aesthetic value but also for their exceptional physical properties. Since ancient times, impurities and defects in diamonds have played a critical role in their appraisal as gemstones. Diamonds exhibit different colors depending on the type and concentration of impurities within the crystal. A yellow hue typically indicates dominant nitrogen impurities (type Ib), whereas a blue color is associated with boron as the primary impurity (type IIb). Two types of colorless diamonds exist: type Ia, which contains nitrogen in dimer form, and type IIa, which contains only trace impurities.

In recent years, the development of synthetic diamond growth techniques has attracted attention not only for semiconductor applications~\cite{Kasu2021}, but also for exploring quantum properties of defects and their potential technological uses~\cite{Zaitsev2001, Balmer2009, Sumiya2015, Eaaton2016}. In particular, spin states associated with atomic vacancies in diamond are regarded as promising candidates for solid-state quantum sensors, with applications ranging from life sciences to materials research~\cite{Acosta2013, Glenn2015, Morishita2019, Barry2020}. These spin states also offer a platform for fundamental studies in quantum physics and pave the way toward quantum technologies, including quantum computation and communication~\cite{Hensen2015, Tashima2019, Cujia2019, Childress2013, Nakazato2022, Rozpedek2019}. Most diamond-based quantum technologies to date have relied on commercially available synthetic diamonds fabricated via high-pressure high-temperature (HPHT) or chemical vapor deposition (CVD) methods.

More recently, significant advances in diamond synthesis technologies have enabled the growth of large, high-quality single crystals. The HPHT flux and CVD methods are now commonly employed to produce type-IIa diamonds with extremely low vacancy concentrations. However, the formation of single-atom vacancies at the sub-ppb level is thermodynamically unavoidable during the growth process~\cite{SM} and must be carefully controlled for quantum technology applications. Previous studies have primarily reported ppm-level concentrations of neutral single vacancies (V$_0$) [Fig. 1(a)], inferred from zero-phonon lines of optical centers attributed to V$_0$~\cite{Bogdanov2018, Stoneham1977, Subedi2021}.

Although numerous studies have explored the engineering of nitrogen-vacancy (NV) color centers  [Fig. 1(c) middle]—defect states formed by radiation-induced atomic vacancies adjacent to nitrogen atoms~\cite{Pezzagna2011, Lhemann2018, Sumikura2020, Luo2022}—only limited attention has been given to the intrinsic structure and quantum properties of defect centers in non-irradiated diamonds, owing to the extremely low concentration of vacancies. In particular, the quantum ground state of single atomic vacancies in non-irradiated synthetic diamonds, which are widely used as substrates for quantum devices, remains largely unexplored. Here, the term “quantum ground state” encompasses not only the spin state, but also the many-body ground state arising from electron-phonon coupling at the vacancy site. Understanding these intrinsic quantum states at cryogenic temperatures is essential for the rational development of next-generation quantum technologies.

Here, we report a pronounced softening of the transverse elastic stiffness constant, $C_{44}$, revealed by ultrasonic measurements below 1 K. This elastic softening may arise from electric quadrupolar degrees of freedom, indicating the presence of an as-yet-unidentified, defect-derived non-magnetic quantum ground state with $T_1$ or $T_2$ symmetry at ppb-level concentrations.

Diamond has a cubic crystal structure (O$_{\rm h}^7$, $Fd\bar{3}m$, No. 227) with lattice constant $a = 3.567 $\AA ~\cite{Rikanenpyou}, as schematically shown in Fig. 1(a). The $sp^3$ orbitals of carbon with the outermost electron configuration $(2s)^2(2p)^2$ are covalently bound to each other, and thereby, form the diamond structure, resulting in one of the hardest materials on earth.  We used three non-irradiated commercially available synthetic diamond species grown by HPHT synthesis (Sample 1 and 3) and CVD (Sample 2) for our ultrasonic measurements presented here.

Sample 1 is a 0.16 carat mono-sectorial type-IIa (colorless) HPHT diamond (New Diamond Technology), with a length of 3.043 mm along the [110] direction and a thickness of $\sim 1.0$ mm along the [001] direction. According to the manufacturer, the B and N concentrations in this type-IIa HPHT diamond without irradiation process are expected to be $\sim$50 ppb and $\sim$10 ppb, respectively. Sample 2 is a type-IIa CVD-grown diamond single crystal, so called `Electronic Grade' (Element Six) without irradiation, which has a B concentration of less than 1 ppb and a N concentration of 0.1-1 ppb, and natural abundance of $^{13}$C impurity (98.892 : 1.108 = $^{12}$C : $^{13}$C). The dimension of the sample is $2.060 \times 2.18 \times 0.504$ mm$^3$. Sample 3 is a non-irradiated  type-Ib (yellow) HPHT diamond (Sumitomo Electric), trade name SUMICRYSTAL (UP $3 \times 3 \times 2$ 100c), with a length of 2.944 mm along the [001] direction and a thickness of $\sim 2.0$ mm along the [100] direction.

Ultrasound was generated and detected by using a pair of LiNbO$_3$ resonance transducers of 100 $\mu$m thickness (with fundamental frequency of $\sim18$ MHz), which are bonded on the polished sample surfaces with room-temperature-vulcanizing silicone. We measured the elastic constant $C_{44}$ of sample 1 using the transverse ultrasonic wave propagating along [110] with polarization along the [001] axis, which induced the elastic strain $\frac{1}{\sqrt{2}}(\varepsilon_{\rm yz}+\varepsilon_{\rm zx})$. Similarly, we measured $C_{44}$ in samples 2 and 3 using the transverse wave propagating along [001] with polarization along the [100] axis, which induced the elastic strain $\varepsilon_{\rm zx}$. We converted the sound velocity $v_{\rm ij}$ to the elastic constant $C_{\rm ij}$ using the formula $C_{\rm ij} = \rho v_{\rm ij}^2$,~\cite{Luethi2006} where $\rho = 3.515$ g/cm$^3$ is the calculated density of diamond. We calculated the absolute value using the measured sound velocity of 12733 m/s at 1 K.

For our low-temperature ultrasonic measurements, we used two different $^3$He-$^4$He dilution refrigerators: top-loading, wet- type for Samples 1 (0.025-0.65 K) and 3 (0.025-0.65 K), and dry-type for Sample 2 (0.025-2.0 K), and $^3$He refrigerators for Samples 1 (0.5-2.0 K). Measurements above 2 K were made using PPMS (Quantum Design Inc.). Magnetic fields up to 16.5 T were generated by superconducting magnets.

\begin{figure}
\begin{center}
\includegraphics[width=0.95\linewidth]{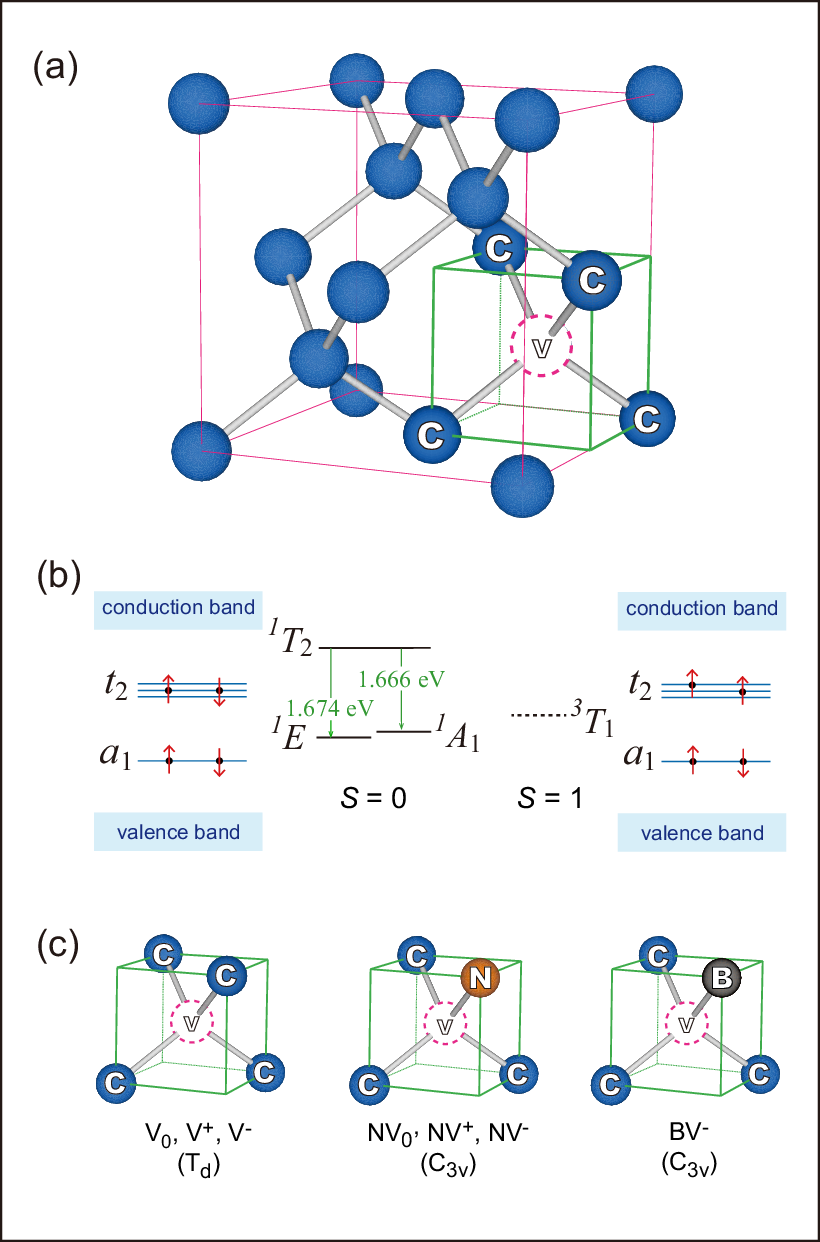}
\caption{(a) Diamond crystal structure with a single atomic vacancy in a unit cell. (b) Electron occupation of the ground-state moleculear orbitals configuration~\cite{Coulson1957}, and energy level structure of V$_0$~\cite{Davies1981}.  The red arrows represent electron spins and their arrangement. The optically inactive $^3T_1$ state of V$_0$ is indicated by dashed line. (c) Schematic illustrations of common defects with atomic vacancies in non-irradiated diamond; (from left) single atomic vacancies, NV centers (a nitrogen atom adjacent to a vacancy), BV$^-$ center (a boron atom adjacent to a vacancy). The symmetry of each defect is also represented.}
\end{center}
\end{figure}

The temperature dependence of the elastic constant $C_{44}$ at zero magnetic field is shown in Figs. 2(a) and 2(c) for single-crystalline HPHT type-IIa diamond (Sample 1). $C_{44}$ gradually hardens down to $\sim$1 K [as shown in Fig. 2(c)], which is a general behavior of solids owing to the anharmonicity of acoustic phonons. This change is consistent with previously reported temperature dependences of the sound velocity of transverse ultrasonic waves measured down to $\sim$10 K~\cite{McSkimin1972, Migliori2008, Nagakubo2016}. To the best of our knowledge, no previous work has yet studied the elastic constants of diamond below 1 K. The low-temperature region of $C_{44}$ exhibits 125 ppm softening from 1 K down to 20 mK. These results indicate that the sites causing elastic softening locally preserve the tetrahedral $T_{\rm d}$ symmetry even at 20 mK. Figure 2(b) shows the magnetic-field dependence of $C_{44}$ at 25 mK and 2 K with shifted offset for the 2 K data. The magnetic field is applied along the [001] direction. The 125 ppm softening is almost unchanged for magnetic fields up to 16.5 T.

\begin{figure}[t]
\begin{center}
\includegraphics[width=0.9\linewidth]{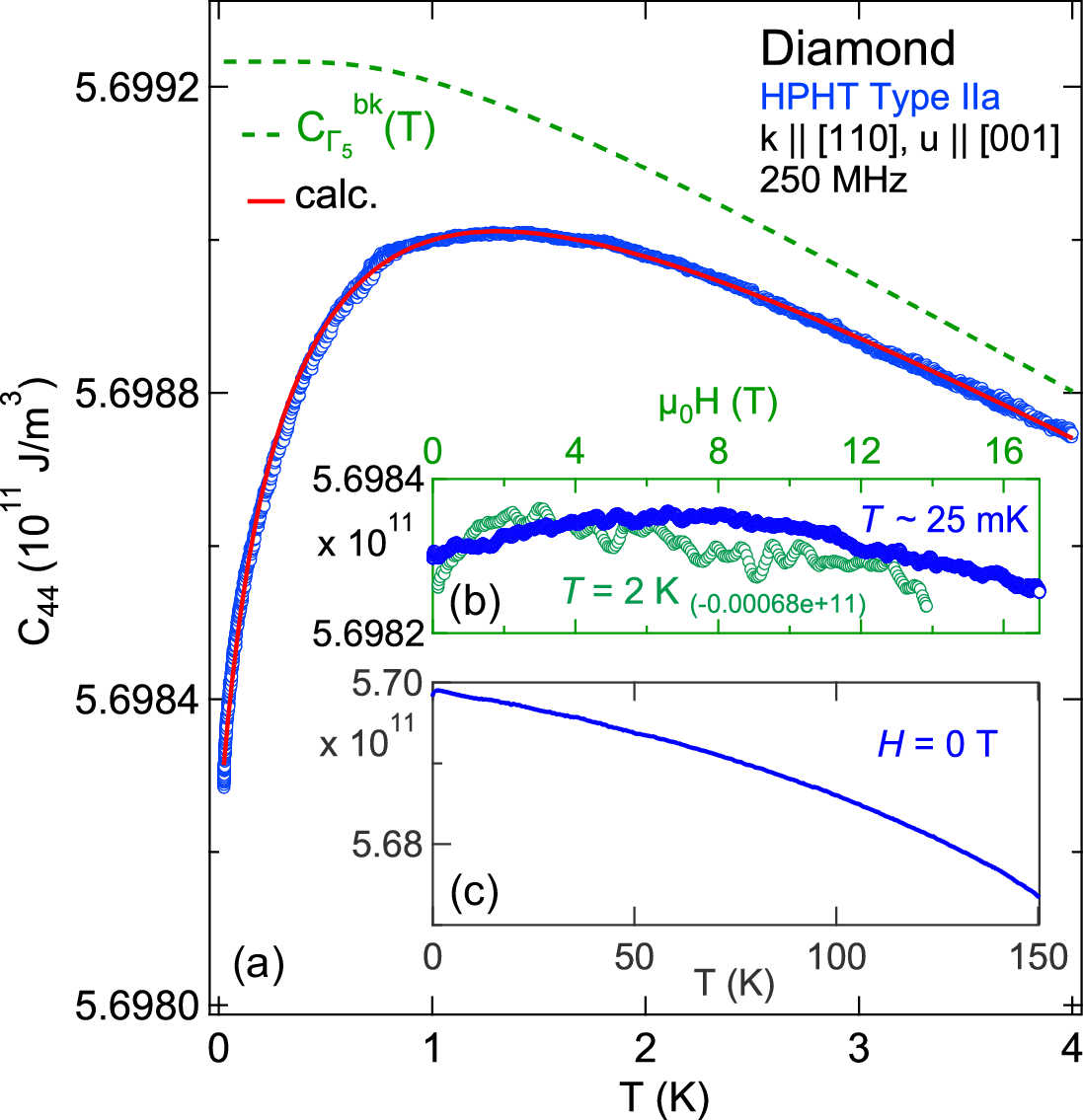}
\caption{\label{fig:fig2} (a) Temperature dependence of the elastic constant $C_{44}$ for single-crystalline synthetic HPHT diamond (type IIa) below 4 K at zero magnetic field compared to calculations (solid red curve) based on the orbital analogue of the Curie-Weiss law (see text). The green dashed line represents the background term $C_{\Gamma 5}^{bk}$ owing to phonons uncoupled to the electronic system. (b) Magnetic-field dependence of $C_{44}$ at 25 mK (blue) and 2 K (green) with shifted offset for the 2 K data. The magnetic field is applied along the [001] direction. (c) Temperature dependence of the elastic constant $C_{44}$ for a wide temperature region up to 150 K.}
\end{center}
\end{figure}

Further, we investigated the elastic response for other diamonds grown by different methods. Figure 3 shows the temperature dependence of $C_{44}$ for three different types of diamonds. The inset of Fig. 3 shows data up to 100 K on a logarithmic temperature axis. Sample 1 is the type-IIa HPHT diamond with data already shown in Fig. 2. Sample 2 is also a type-IIa diamond grown by CVD. Sample 3 is a HPHT yellow diamond of type Ib that presumably contains numerous nitrogen impurities. Sample 1 and 2 show a nearly identical low-temperature softening within the resolution of the present measurements. The data for Sample 2 are noisy because of the small sample thickness. The softening in Sample 3 (type-Ib HPHT) is only approximately 1/3 of that of the Samples 1 (type-IIa HPHT) and 2 (type-IIa CVD). Obviously, the defect concentration responsible for the softening in sample 3 is lower than in samples 1 and 2.  Notably, the  low-temperature softening shows nearly zero magnetic-field dependence up to 14.0 - 16.5 T  (Fig. A2 in [\onlinecite{SM}]). These results simply suggest that the quantum ground state, which is the origin of the softening, is non-magnetic.

\begin{figure}[t]
\begin{center}
\includegraphics[width=0.95\linewidth]{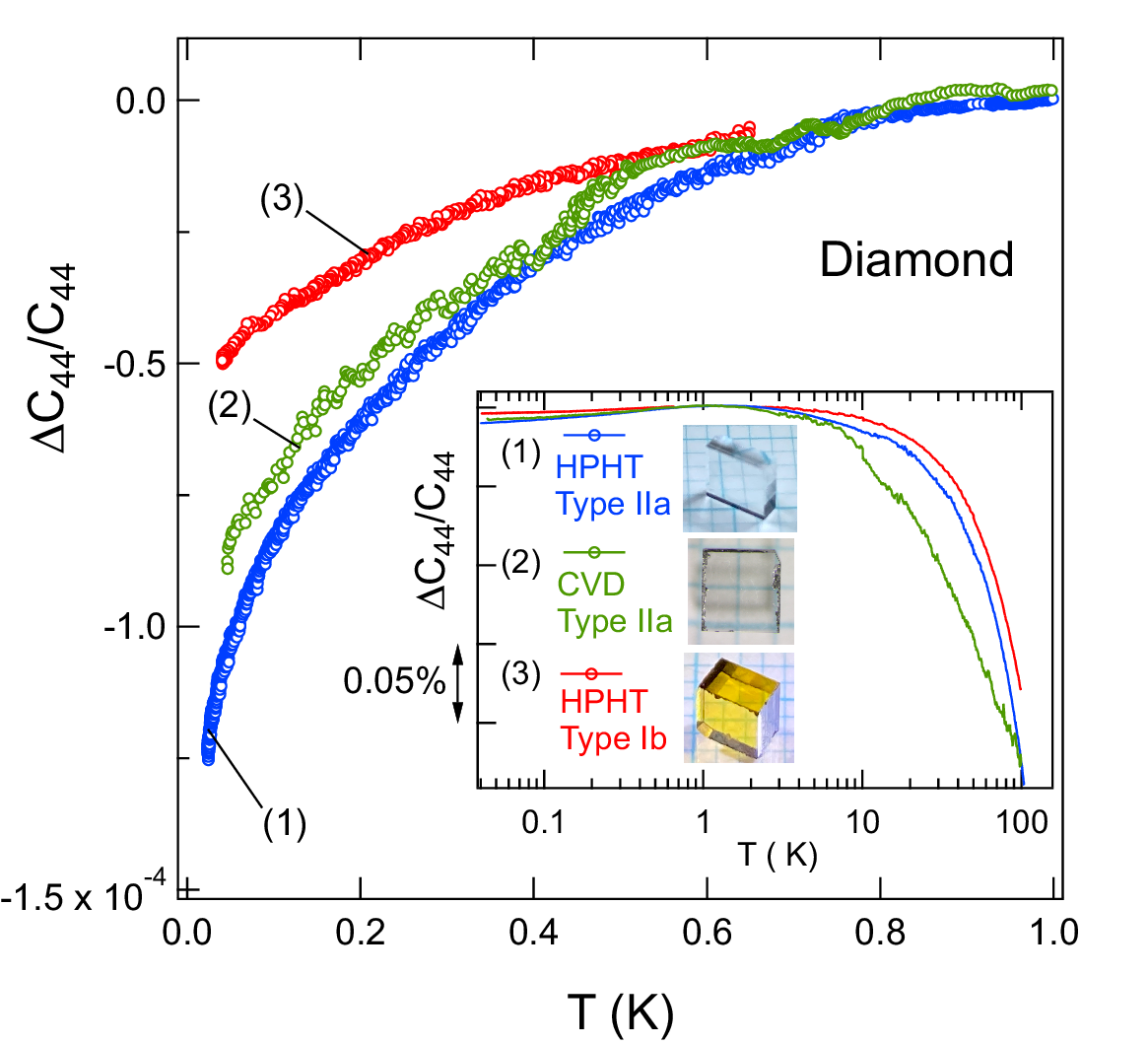}
\caption{\label{fig:fig3}Comparison of the elastic softening of $C_{44}$ for three different types of diamond. (1: blue symbols) HPHT type IIa (see also Fig. 2), (2: green symbols) CVD type IIa, and (3: red symbols) HPHT type Ib. The relative variation in $\Delta C_{44}/C_{44}$ was normalized at 1 K (The amount of change in Sample 3 was extrapolated from 0.65 K to 1 K). The inset shows results up to 100 K in a semilogarithmic scale. Photographs of the samples used for the measurements are shown as well.}
\end{center}
\end{figure}

The low-temperature softening we found in diamond (as shown in Figs. 2 and 3) is reminiscent of the elastic softening found by Goto {\it et al.} in their previous ultrasound investigations on Si, where they successfully estimated ppb levels of V$_0$ vacancy concentrations in an undoped Si crystal and ppb levels of V$^+$ in B-doped Si grown by a floating-zone method~\cite{Yamada-Kaneta2006, Goto2006, Goto2015, Okabe2013, Mitsumoto2014, Baba2011, Goto2007}. In non-doped Si, the neutral vacancy $V_0$ has been proposed to have a $^3T_1$ ground state due to strong inter-orbital coupling.~\cite{Matsuura2008} In analogy to the interpretation for the similar softening observed in Si, it is likely that some molecular orbitals in diamond maintain a degenerate electronic state down to low temperatures, {\it i.e.}, the Curie-type softening of the transverse $C_{44}$ mode (corresponding to the electric quadrupolar susceptibility of $\Gamma_5 (T_2)$ symmetry~\cite{SM}) in diamonds could also be connected to the $T_1$ or $T_2$ state in $T_{\rm d}$ point group. 

However, the elastic softening mechanism proposed for Si [\onlinecite{Goto2006}] is unlikely to apply to diamond. In contrast, for $V_0$ in diamond, the electric ground state is believed to have vibronic many-body quantum states $^1E$ and $^1A_1$ with a gap of 8 meV (Fig. 1(b), middle) from the various experiments on irradiated diamonds~\cite{Davies1994}. Recent ab-initio calculations~\cite{Prentice2017} also demonstrate that the dynamical Jahn-Teller-distorted structure of $T_{\rm d}$ symmetry is lower in energy than the static Jahn-Teller distorted structure of $D_{\rm 2d}$~\cite{Clark1973}, in agreement with experimental observations. Considering only the local symmetry of the low-lying vibronic quantum ground states $^1E$ and $^1A_1$, these states lack $T_2$-type electric quadrupolar degrees of freedom and are therefore unlikely to couple directly to the $T_2$-symmetric phonon mode associated with $C_{44}$, {\it i.e.,} the doubly degenerate spin-singlet state with $E$ symmetry cannot lead to the softening of the elastic constant $C_{44}$.

The alternative electronic state, $^3T_1$, associated with $V_0$ in diamond (Fig. 1(b), right), is optically inactive and has consequently been less extensively investigated. This state possesses $\Gamma_5$ ($T_2$) electric quadrupolar degrees of freedom; however, it is theoretically predicted, based on perturbative calculations, to lie approximately 0.2 eV above the $^1E$ ground state~\cite{Wyk1995}. Although it may be possible that the optically inactive $^3T_1$ state of $V_0$ constitute the ground state due to strong inter-orbital coupling, analogus to the case in Si~\cite{Matsuura2008}, this interpretation remains speculative. This theoretical framework for Si also predicts that the elastic softening persists under magnetic fields on the order of 10 T, except at ultra-low temperatures, owing to the minimal lifting of orbital degeneracy via spin-orbit interaction and the orbital Zeeman effect. This is consistent with the observation that the softening in diamond, as obtained in the present study, insensitively to applied magnetic fields. (Fig. A2 in [\onlinecite{SM}]) Nevertheless, the ultrasonic measurements presented here do not provide definitive evidence supporting $^3T_1$ as the ground state. Therefore, guided by prior theoretical insights, we conclude that the observed softening in diamond likely originates from defect states distinct from the neutral vacancy $V_0$. Further experiments, such as measurements of different ultrasonic modes, investigations under higher magnetic fields, and studies with varying magnetic field orientations, are required to rigorously evaluate the potential involvement of the $^3T_1$ state.

\begin{table}
\caption{Established electronic states of vacancy centers in diamond~\cite{Davies1981,Mainwood1997}}
\label{t1}
\begin{center}
\begin{tabular}{lccccr}
\hline
\textrm{Structure}&
\textrm{Point}&
\textrm{Electronic}&
\textrm{Ground}&
\textrm{Energy}&
\textrm{Refs.}
\\
\textrm{}&
\textrm{Group}&
\textrm{Config.}&
\textrm{State}&
\textrm{Gap (eV)}&
\textrm{}
\\
\hline
V$^-$ 	& $T_{\rm d}$ 	& $(a_1)^2 (t_2)^3$ 	& $^4A_2$ 		&3.150	&[\onlinecite{Davies1992}]\\
V$_0$ 	& $T_{\rm d}$ 	& $(a_1)^2 (t_2)^2$ 	& $^1E$ 			&1.674	&[\onlinecite{Davies1992}]\\
V$^+$ 	& $T_{\rm d}$ 	& $(a_1)^2 (t_2)^1$ 	& $^2T_2$ 		&$\sim$2.8$^{\S}$		&[\onlinecite{Coulson1971}]\\
NV$^-$ 	& $C_{\rm 3v}$ 	& $(a_1)^2 (e)^2$ 	& $^3A_2$ 	&1.945	&[\onlinecite{Davies1992}]\\
NV$_0$ 	& $C_{\rm 3v}$ 	& $(a_1)^2 (e)^1$ 	& $^2E$ 		&2.156	&[\onlinecite{Siyushev2013}]\\
NV$^+$ 	& $C_{\rm 3v}$ 	& $(a_1)^2$ 		& $^1A_1$ 	&-		&[\onlinecite{Meara2019}]\\
BV$^-$ 	& $C_{\rm 3v}$ 	& $(a_1)^2(e)^2$ 	& $^3A_2$ 	&-		&[\onlinecite{Umeda2022}]\\
\hline
\end{tabular}
\end{center}
\hspace{5 mm} \footnotesize{$^{\S}$ theory~\cite{Mainwood1997}}
\end{table}
In the following, we examine whether other types of known defects (as shown in Fig. 1(c)) in non-irradiated diamond could account for the observed softening. Some of the electronic ground states of the various vacancy centers in diamond are listed in Table 1. Except for the neutral vacancy V$_0$, only the negatively charged vacancy V$^-$, and positively charged vacancy V$^+$ preserve $T_{\rm d}$ symmetry~\cite{Coulson1971}. The possibility of V$^+$ and V$^-$ can also be ruled out for the following reasons. Stable existence of V$^+$ and V$^-$ in the crystal requires electron donors and acceptors to V$_0$, which are generally provided by N and B doping, respectively. If they were the cause of the softening, it would be strongly dependent on sample, with different B and N concentrations. On the other hand, the three diamonds measured in this study have N and B concentrations that differ by an order of magnitude, but the magnitude of the softening varies much less. This fact strongly suggests that V$^+$ and V$^-$, as well as P$_1$ (N substitutional), B$_s$ (B substitutional), NV and BV (N and B complex) color centers~\cite{Cox1994}, are not the origin of the softening.

As shown in Table 1, the NV centers~\cite{Cox1994, Doherty2013} possess local $C_{3v}$ symmetry due to Jahn-Teller distortion and therefore do not contribute to the elastic softening~\cite{SM}. This is also true for the P$_1$ centers, N$_n$V clusters ($n$ = 2-4; substitutional nitrogen atoms surrounding a vacancy), divacancies (V$_2$, consisting of two adjacent missing carbon atoms), large multi-atom vacancy V$_n$ clusters~\cite{Kamihara2022},  and hydrogen-related complexes and other defects involving local symmetry breaking due to an asymmetric arrangement of atoms. Interactions of these defects with phonons will appear via spin-orbit coupling, which is negligible in diamond~\cite{Lenef1996}. Our experimental finding that the softening is insensitive to magnetic fields further hints that nonmagnetic ground states are responsible for the softening, {\it i.e.}, magnetic ground states ({\it e.g.}, V$^-$, V$^+$, NV$^-$, NV$_0$, and BV$^-$) can be excluded as candidates. However, as mentioned above, depending on the strength of the spin-orbit interactions~\cite{Matsuura2008}, it remains possible that even for a magnetic state such as spin-triplet $^3T_1$, the magnetic field response could be sufficiently weak within the measured field and temperature ranges. Thus, we cannot definitively rule out all magnetic ground states with finite spin.

Although the known defects discussed above have been ruled out as the origin of the softening, non-irradiated diamond may still host as-yet-unidentified vacancies or interstitials that preserve local $T_d$ symmetry. Their quantum ground states likely remain optically inactive or undetermined, and could account for the observed elastic softening. The possible physical picture envisaged here is a situation where the orbitally degenerated multi-electronic state of $T_1$ or $T_2$ is coupled to the $T_2$-phonon mode, and forming a vibronic state due to the dynamical Jahn-Teller effect. It will allow the system to preserve tetrahedral symmetry without causing static Jahn-Teller distortion, similar to the quantum tunneling state in the cubic clathrate crystals~\cite{Goto2004}. Another example is the cubic, non-Kramers $\Gamma_3$ ground state system of PrMg$_3$, which has no long-range order down to 50 mK, where a quantum-mechanical hybridization of $4f$ electrons and phonons is believed to form the vibronic state.~\cite{Araki2012}. To clarify the origin of the softening, it would be desirable to compare irradiated and non-irradiated samples, and to perform similar measurements on samples with controlled impurity profiles—including dislocations and elongation defects—while systematically varying the concentrations of nitrogen, boron, carbon vacancies, and other relevant impurities. This represents a challenge for future investigations. 

Finally, we follow the phenomenological analysis of quadrupolar susceptibility. The solid red line in Fig. 2(a) is the result of calculations based on the equation ${C_{\Gamma 5} = C_{\Gamma 5}^{bk}(T-T_{\rm C})/(T-\Theta)}$ (for details, see [\onlinecite{SM}]). The fit yields $T_{\rm C} = -260.196$ mK,  $\Theta = -260.242$ mK, and  the Jahn-Teller energy $\Delta_{\rm JT} = T_{\rm C} - \Theta = 0.046$ mK. The dashed line represents the phenomenological fit for the background contribution ${C_{\Gamma 5}^{bk} = C_{\Gamma 5}^{0}-s/\{\exp(t/T)-1\}}$~\cite{Luethi2006}, with $C_{\Gamma 5}^{0} = 5.69923 \times 10^{11}$ Jm$^{-3}$, $s = -4.681226 \times 10^7$ Jm$^{-3}$, and $t$ = 2.941652 K. The negative value of $\Theta$ indicates antiferro-type inter-site interactions for the electric quadrupole on the unspecified defect. The negative value of $T_{\rm C}$ suggests quadrupolar fluctuations associated with the degenerate ground state even at the lowest temperature of $\sim 20$ mK. This result further proves that the cubic site symmetry $T_{\rm d}$ at the site of defects that cause softening is preserved in the investigated diamond material, {\it i.e}., the local distortion owing to the static Jahn-Teller effect is irrelevant for diamond with low vacancy and impurity concentrations. 
If we roughly estimate the concentration of the unspecified defects which have electric quadrupole degrees of freedom using the relation $N = \Delta_{\rm JT}C_{\Gamma}^0/(\delta_{\Gamma}^0)^2$ for the acoustic mode $C_{44}$, which is assuming a $T_2$-triplet ground state and was used to analyze the vacancy concentration in Si~\cite{Goto2006}. Here, $N$ denotes the concentration and $\delta_{\Gamma}^0$ the deformation binding energy of unspecified defect. As a rough estimation, we assume the formation energy 9-15 eV of the vacancies in diamond~\cite{Bourgoin1983}, then we obtain the concentration $N \sim 0.98-0.36$ ppb. If the unspecified defects possess similar formation energies, this would imply that all diamonds examined in this study host defects with electric quadrupole moments on the order of ppb.

In summary, we performed ultrasonic measurements down to 20 mK on non-irradiated synthetic diamonds and discovered anomalous elastic softening. Although the origin of this softening remains unresolved, this result suggests the presence of an unspecified defect-derived quantum ground state with $T_1$ or $T_2$ symmetry in synthetic diamonds. Our phenomenological analysis implies that the concentration of the defect is at the ppb level. The present results are particularly relevant for researchers employing synthetic diamonds in diverse fields, including quantum information, biological sensing, and power electronics. Moreover, our findings offer a new avenue for exploring the quantum ground states of vacancies in diamond. Further studies of the elastic response of various diamond types at cryogenic temperatures are essential to elucidate the underlying physics of diamond vacancies and defects, with potential implications for future technological applications.\\

\begin{acknowledgements}
We thank Profs. Terutaka Goto, Takashi Taniguchi, Hiroaki Kusunose, and Satoru Hayami for helpful discussions. The present research was supported by JSPS KAKENHI Grants Nos. JP23H04868, JP21KK0046, JP22K03501, and Toyota Physical and Chemical Research Institute under the 2021 Toyota Riken Scholar Collaborative Research Program (Phase 1) to TY and TT. We acknowledge support from the DFG through the W\"urzburg-Dresden Cluster of Excellence on Complexity and Topology in Quantum Matter - {\it ct.qmat} (EXC 2147, project-id 390858490) and from the HLD at HZDR, a member of the European Magnetic Field Laboratory (EMFL). TY would like to acknowledge Mr. Kousuke Nakamura and Mr. Tatsuji Meike at Hokkaido University Technical Support Division for assistance in polishing the diamonds. TY and RH would like to thank Prof. Atsuhiko Miyata for supporting the measurements at HZDR.
\end{acknowledgements}

\typeout{}
\bibliography{yanagisawa_Diamond2025.bib}

\renewcommand{\thesection}{\Roman{section}}.
\renewcommand{\thetable}{A\Roman{table}}
\renewcommand{\thefigure}{S\arabic{figure}}
\setcounter{table}{1}
\setcounter{figure}{0}
\onecolumngrid
\begin{center}
\vspace{5mm}
\newpage
{\bf \large Supplemental Material for\\ 
Elastic Softening in Synthetic Diamonds}
\vspace{5mm}

Tatsuya Yanagisawa, Ruo Hibino, Hiroyuki Hidaka, Hiroshi Amitsuka, Toshiyuki Tashima, Mitsuhiro Akatsu, Yuichi Nemoto, Sergei Zherlitsyn, and Joachim Wosnitza\\
\vspace{5mm}

\end{center}
\onecolumngrid
\section*{\label{sec:Schematic}1. THERMODYNAMIC CONSIDERATIONS OF THE FORMATION OF ATOMIC VACANCIES}

 Point defects are always formed during crystal growth at high temperatures. In other words, it is thermodynamically inevitable that such structural instability exists even in thermal equilibrium because there is always finite entropy at a finite temperature, according to the laws of thermodynamics. If $N$ point defects (single-electron vacancies) form in a crystal consisting of $N_0$ atoms and assuming that $N \ll N_0$ and the formation energy of a single atomic vacancy is $E_{\rm V}$, then Gibbs's free energy is written by,
 \begin{equation}
\Delta G = \Delta U - T\Delta S = NE_{\rm V} - T\Delta S,
 \end{equation}
where the entropy is $S = -k_B\ln W$, with the number of states $\textstyle W = \frac{N!}{(N_0 - N)N!} $. 

From the approximation by Stirling's formula with $N \gg 1$, the entropy can be rewritten as follows,
 \begin{equation}
\Delta S = -k_{\rm B} N \left \{ \frac{N_0}{N} \ln \frac{N_0}{N}-\frac{(N_0-N)}{N} \ln \frac{(N_0-N)}{N} \right \}.
 \end{equation}
Substituting this for $\textstyle \frac{dS}{dE} = \frac{1}{T} $,we obtain the temperature dependence of the number of atomic vacancies
\begin{math}N = N_0 e^{(-E_{\rm V}/T)} \end{math}. Although the atomic vacancy-formation energy is unknown, this equation shows that the number of atomic vacancies increases rapidly during crystal growth with increasing temperatures.
\vspace{10mm}
 
 \section*{\label{sec:Schematic}2. ULTRASONIC METHOD}
  \begin{figure}[h]
\centering{
TABLE AI:~~Symmetry, elastic constant, symmetrized strain and quadrupole, and illustration of strain and coupled charge distributions.\\}
 \vspace{10mm}
\includegraphics[width=0.7\linewidth]{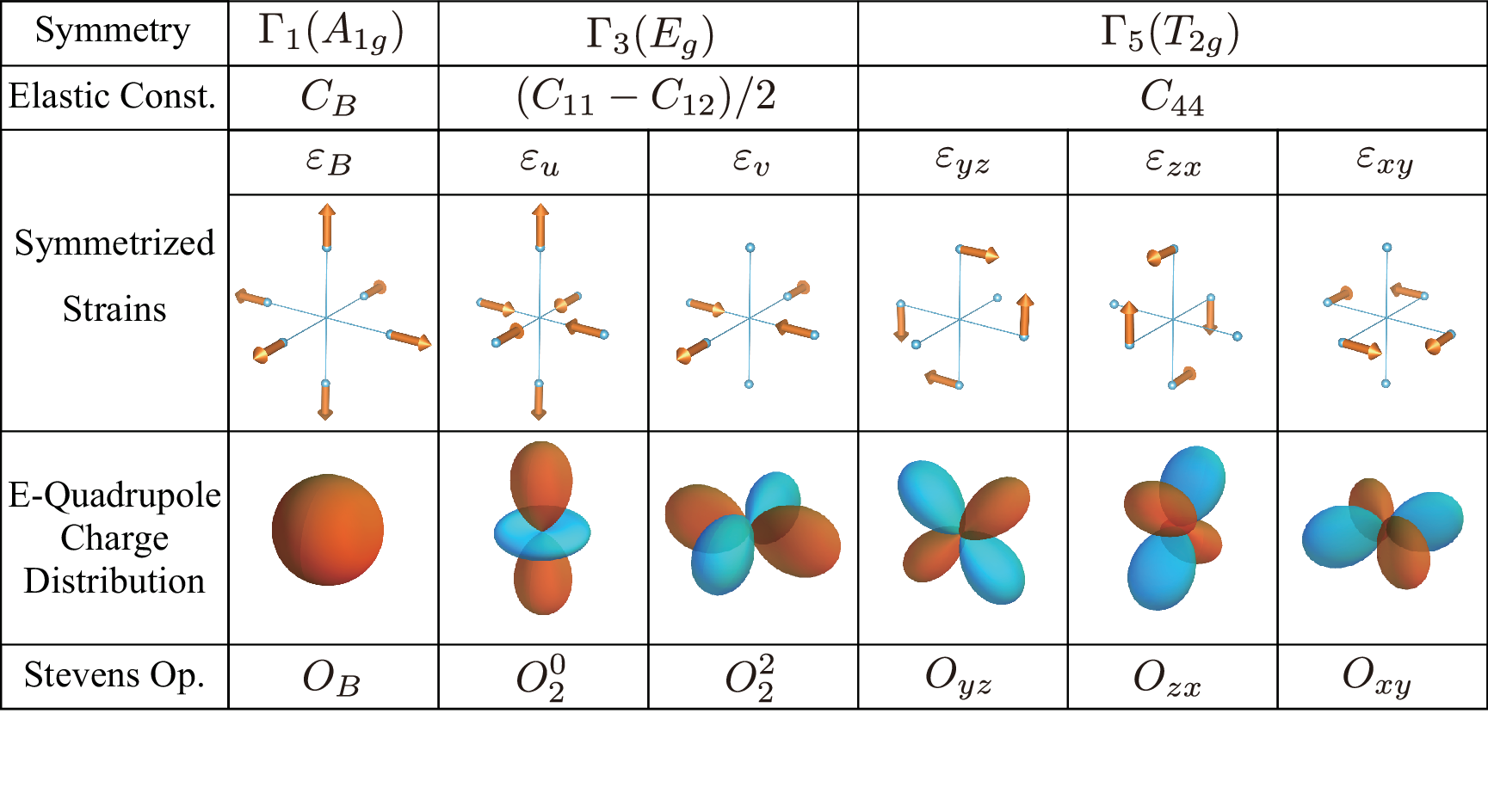}
\end{figure}

 Ultrasonic measurements have been widely used in studies of solid-state physics. They are valuable tools for detailed characterization of the lattice elasticity as well as any other type of phenomenon that couple to the strain fields. Some examples are phenomena involving electric quadrupolar degrees of freedom, magnetic ordering (via exchange-striction coupling), superconductivity, local Einstein phonons, magneto-acoustic quantum oscillations, and so on ~\cite{Luethi2006}. By using longitudinal and transverse ultrasonic modes, the response of these phenomena can be obtained spectroscopically. In particular, the temperature dependence of the elastic constant is one of the powerful tools to investigate the quadrupolar degrees of freedom in the $sp^3$ dangling bond of the single neutral vacancy V$_0$ in Si and diamond. The $\Gamma_5$-type symmetrized elastic strains $\epsilon_{yz}, \epsilon_{zx}$, and $\epsilon_{xy}$, that correspond to the $C_{44}$ transverse ultrasonic modes, couple to the electric quadrupole with the same symmetry as $O_{yz}, O_{zx}$, and $O_{xy}$, which are active in the quantum ground state.

\newpage
  \section*{\label{sec:Schematic}3.  WAVEFUNCTIONS FOR $sp^3$ HYBRID ORBITALS AND ACTIVE ELECTRIC QUADRUPOLES}

Diamond is covalently bonded by the four outer-shell carbon $2s^12p^3$ electrons, forming hybridized orbitals. As shown in Fig. A1, when the dangling bonds of carbon atoms in the direct neighborhood of a single-atom vacancy are numbered from 1 to 4, the wave functions of the dangling bonds are given by,

\begin{eqnarray}
\phi_1 = \frac{1}{2}(s^{(1)}+p_x^{(1)}+p_y^{(1)}+p_z^{(1)}),\nonumber \\
\phi_2 = \frac{1}{2}(s^{(1)}+p_x^{(1)}-p_y^{(1)}-p_z^{(1)}),\nonumber \\
\phi_3 = \frac{1}{2}(s^{(1)}-p_x^{(1)}+p_y^{(1)}-p_z^{(1)}),\nonumber \\
\phi_4 = \frac{1}{2}(s^{(1)}+p_x^{(1)}-p_y^{(1)}+p_z^{(1)}).
\end{eqnarray}

\begin{figure}[t]
\hspace{-35mm}\includegraphics[width=0.6\linewidth]{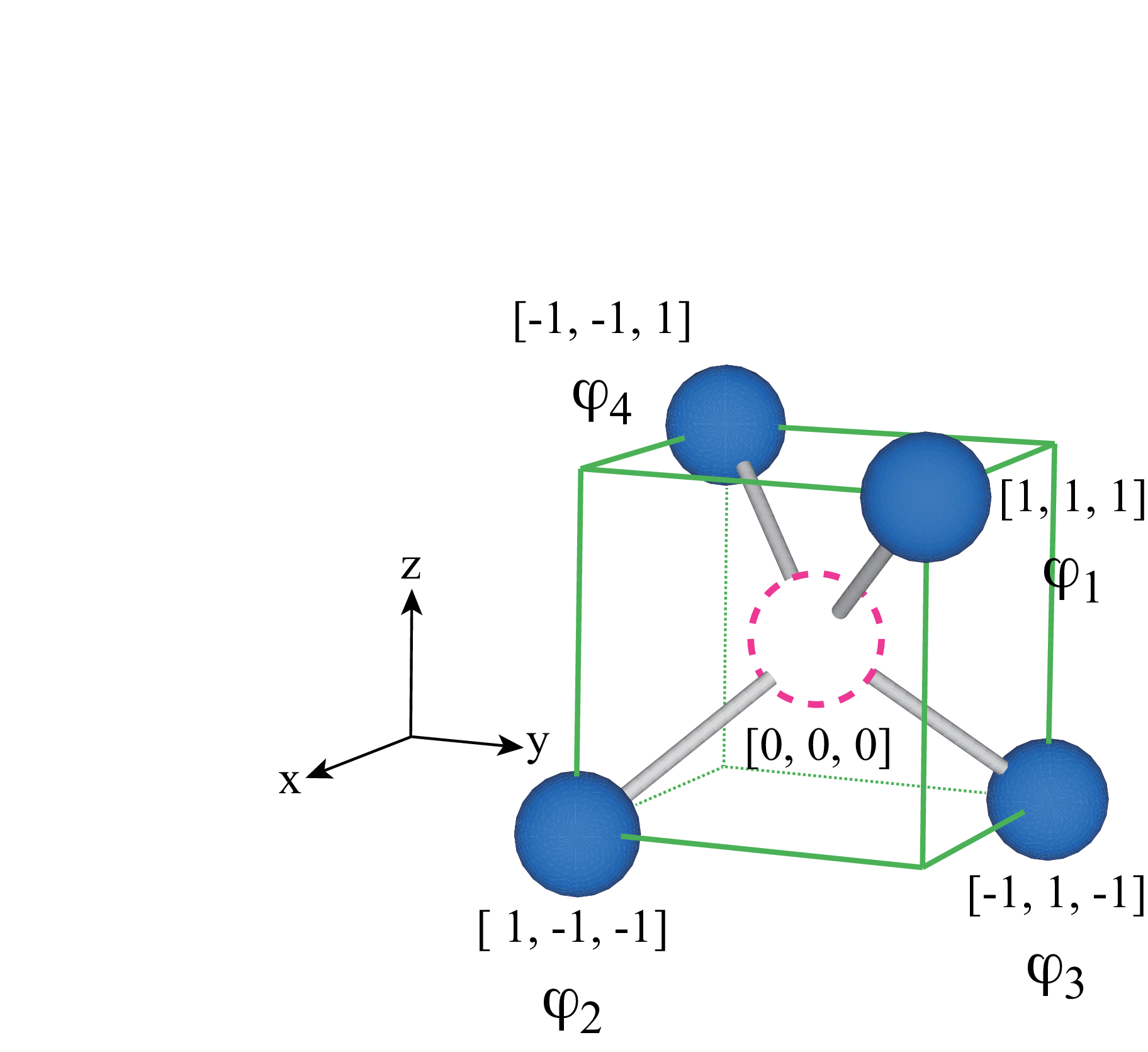}\\
{
FIG. A1~~Scheme of the dangling bonds of a single atomic vacancy.}
\end{figure}

These equations represent the four-fold degeneracy of equivalent $sp^3$ hybrid orbitals. The transfer energy among the $sp^3$ dangling bonds is written as, $-\gamma=\langle\phi_j|\mathscr{H}_{0}|\phi_i\rangle$ ($i, j = 1, 2, 3,$ and 4). Here, the Hamiltonian $H_0$ for the transfer motions among the dangling bonds has cubic symmetry. When we assume the energy of the dangling bonds is $E$, the matrix elements of $H_0$  are written as,

\begin{eqnarray}
\mathscr{H}_{0} =~
\begin{matrix}
   ~~~~~~~~	 &~ |\phi_1 \rangle & ~|\phi_2 \rangle & ~|\phi_3 \rangle & ~ |\phi_4 \rangle \\
   \langle \phi_1| &	~E			& ~-\gamma	& ~-\gamma 	& ~-\gamma \\
   \langle \phi_2|  & 	~-\gamma	& ~E 			& ~-\gamma	& ~-\gamma \\
   \langle \phi_3|  & 	~-\gamma	& ~-\gamma	& ~E 			& ~-\gamma \\
   \langle \phi_4|  &	~-\gamma	& ~-\gamma	& ~-\gamma	& ~E \\
\end{matrix}
\end{eqnarray}
 
 By diagonalizing Eq. (4), we obtain\\
 
\begin{eqnarray}
\mathscr{H}_{0} =~
\begin{matrix}
   ~~~~~~~~	 &|\psi^{a1} \rangle & |\psi^{a2} \rangle & |\psi^{a3} \rangle & |\psi^{a4} \rangle \\
   \langle \psi^{a1}| &	E-3\gamma	& ~0			& ~0 			& ~0 \\
   \langle \psi^{a2}|  & 	~0			& ~E+\gamma	& ~0			& ~0 \\
   \langle \psi^{a3}|  & 	~0			& ~0			& ~E+\gamma 	& ~0 \\
   \langle \psi^{a4}|  &	~0			& ~0			& ~0			& ~E+\gamma \\
\end{matrix}
\end{eqnarray}
 
 This shows that the ground state is a singlet and the excited state splits into a triplet with an energy difference of $4 \gamma$ under the site symmetry $T_d$. The single-electron wavefunctions of the eigenstates are
 
 \begin{eqnarray}
a_1:  | \psi^{(a1)}\rangle = | v\rangle =	\frac{1}{2} (|\phi_1 \rangle + |\phi_2 \rangle + |\phi_3 \rangle + |\phi_4 \rangle) \nonumber  \\ 
t_2: \Biggl\{~
\begin{matrix}
 | \psi_a^{(t2)}\rangle =	 | a\rangle =	\frac{1}{2} (|\phi_1 \rangle + |\phi_2 \rangle - |\phi_3 \rangle - |\phi_4 \rangle) \\ 
 | \psi_b^{(t2)}\rangle =	 | b\rangle  = \frac{1}{2} (|\phi_1 \rangle - |\phi_2 \rangle + |\phi_3 \rangle - |\phi_4 \rangle) \\ 
 | \psi_c^{(t2)}\rangle =	 | c\rangle =	\frac{1}{2} (|\phi_1 \rangle - |\phi_2 \rangle - |\phi_3 \rangle + |\phi_4 \rangle) 
\end{matrix}
\end{eqnarray}

We use the notations related to the irreducible representations $A$, $E$, and $T$ for the symmetry of the system (quantum ground state), and labels $a$, $e$, and $t_2$ for the indivisual electron orbitals. The expression for the degrees of freedom that the excited triplet of the dangling orbitals have in the direct product space is

\begin{equation}
(t_2 \otimes t_2) = A \oplus E \oplus T_1 \oplus T_2.
\end{equation}

Goto {\it et al.} consider the orbital part of $T_2$ state in this product space to be the cause of the softening of Si and conduct a phenomenological analysis.~\cite{Goto2006} Actually, the electronic state of the neutral vacancy V$_0$ has two electrons in the excited triplet $t_2$, which is the electron configuration of ($a_1^2t_2^2$). Thus, it is important to consider multi-electron states including Pauli's exclusion rule for spin degrees of freedom, which can be decomposed into~\cite{Coulson1957}
\begin{equation}
^1A_1 \oplus ^1E \oplus ^3T_1 \oplus ^1T_2.
\end{equation}

For neutral vacancy $V_0$ in diamond, it is both theoretically and experimentally confirmed that $^1E$ is the ground state.~\cite{Davies1994, 41}  On the other hand, for the neutral vacancy $V_0$ in non-doped Si, it is proposed from theoretical calculations that $^3T_1$ could be the ground state due to strong inter-orbital interactions.~\cite{Matsuura2008}

The wavefunctions for the each state of the electron configuration of ($a_1^2t_2^2$) are expressed in a simplified form as follows,
\begin{eqnarray}
^1A_1:&(| v \overline{v} a\overline{a}\rangle + | v\overline{v}b\overline{b}\rangle + | v\overline{v}c\overline{c}\rangle )\nonumber  \\ 
^1E:& ( | v\overline{v}a\overline{a}\rangle - | v\overline{v}b\overline{b}\rangle )\nonumber  \\ 
^1T_2:&(| v\overline{v}a\overline{b}\rangle - | v\overline{v}\overline{a}b \rangle) \nonumber  \\ 
^3T_1:& ( | v\overline{v}ab\rangle)
\end{eqnarray}
Here, the wavefunctions are unnormalized. The bars above orbital denote spin states, and only those with $S_z = S$ are shown. For each degenerate representation, only a single component is presented.~\cite{Coulson1957}

The product space of the triplet states are
\begin{equation}
T_1 \otimes T_1 = T_2 \otimes T_2 = A_1 \oplus A_2 \oplus E \oplus T_1 \oplus T_2.
\end{equation}
Thus, both triplet state possess active electric quadrupole with $\Gamma_3 (E)$ and $\Gamma_5 (T_2)$ symmetries.

Here, we introduce the Gell-Mann matrices $\lambda_i$  ($i=1,2, ... ,8$)  of the generator of the rotation group SU(3) representing the active electric quadrupoles with $\Gamma_3 (E)$ and $\Gamma_5 (T_2)$ symmetries as follows,

 \begin{eqnarray}
\Gamma_3 (E):&&\nonumber \\
  O_u &=&  ~\lambda_8 = \frac{1}{\sqrt{3}}
\begin{pmatrix}
	1 & 0 & 0 \\
	0 & 1 & 0 \\
	0 & 0 & -2 \\
\end{pmatrix}~~~~~~~~~~~~~~\nonumber\\
 O_v &=&  ~\lambda_1 = \frac{1}{\sqrt{3}}
\begin{pmatrix}
	0 & 1 & 0 \\
	1 & 0 & 0 \\
	0 & 0 & 0 \\
\end{pmatrix}~~~~~~~~~~~~~~~~~\\
\Gamma_5 (T_2):&&\nonumber \\
  O_{yz} &=&  \frac{1}{\sqrt{2}}(\lambda_5-\lambda_7) = \frac{1}{\sqrt{2}}
\begin{pmatrix}
	0 & 0 & -i \\
	0 & 0 & i \\
	i & -i & 0 \\
\end{pmatrix}\nonumber\\
  O_{zx} &=&  \frac{1}{\sqrt{2}}(\lambda_4+\lambda_6) = \frac{1}{\sqrt{2}}
\begin{pmatrix}
	0 & 0 & 1 \\
	0 & 0 & 1\\
	1 & 1 & 0 \\
\end{pmatrix}~~~~\nonumber\\
  O_{xy} &=&  -\lambda_2 = \frac{1}{\sqrt{2}}
\begin{pmatrix}
	0 & i & 0 \\
	-i & 0 & 0\\
	0 & 0 & 0 \\
\end{pmatrix}~~~~~~~~~~~~~~
\end{eqnarray}

When the local charge fluctuations of the charge state of the vacancy site is denoted by $Q_{\Gamma}$, it is connected to the quadrupole operator obtained above by the relation:

\begin{equation}
Q_{\Gamma} = Z^* a^{*2}O_{\Gamma}.
\end{equation}

Here, $Z^*$ is the effective charge and $a^*$ is the effective radius of the quadrupole with symmetry $\Gamma$.
\vspace{10mm}

  \section*{\label{sec:Schematic}4.  QUADRUPOLAR SUSCEPTIBILITY }

The temperature dependence of the elastic constant in the present paper is calculated using a theory based on the Wigner-Brillouin perturbation method. The crystalline electric field (CEF) Hamiltonian with electric-strain-mediated perturbation is

\begin{equation}
 \mathscr{H}=\mathscr{H}_{0}+\sum_{\Gamma}\frac{\partial \mathscr{H}_{0}}{\partial \epsilon_\Gamma}\epsilon_\Gamma.
 \end{equation}
Here, $\mathscr{H}_{0}$ is derived from the electrostatic potential in the absence of strain $\epsilon_{\Gamma}$. The second term in Eq. (14) is given in terms of electric quadrupole-strain interaction. To describe the elastic softening in $C_{44}$, we adopt the quadrupole-strain interaction regarding $\Gamma_3$ and $\Gamma_5$ symmetry as

\begin{eqnarray}
\mathscr{H}_{\rm QS}&=&  -g_{\Gamma_3}O_{\Gamma_3}\epsilon_{\Gamma_3}-g_{\Gamma_5}O_{\Gamma_5}\epsilon_{\Gamma_5}  \nonumber\\
&=& -g_{\Gamma_3}(O_{u}\epsilon_{u}+O_{v}\epsilon_v)-g_{\Gamma_5}(O_{yz}\epsilon_{yz}+O_{zx}\epsilon_{zx}+O_{xy}\epsilon_{xy}),
\end{eqnarray}

where $g_{\Gamma}$ is a coupling constants, and $O_{\Gamma}$ is an electric quadrupolar moment (as listed in Table AI).\\

Here, $g_{\Gamma_3}$ and   $g_{\Gamma_5}$ are coupling constant. $O_{\Gamma_3}$ and $O_{\Gamma_5}$  are the electric quadrupoles belonging to the $\Gamma_3$ and $\Gamma_5$ symmetries, respectively, as defined in the previous section. In the ultrasonic experiments, the symmetrized strain $\epsilon_\Gamma$ is induced by ultrasound. The quadrupole-strain interaction is incorporated as a perturbed crystal-field level $E_i (\epsilon_\Gamma)$ as a function of symmetrized strain, and we consider strain up to second-order perturbation. The crystal-field energy can be written as:

\begin{eqnarray}
E_i(\epsilon_\Gamma)&=&E_i^0-\left<i|\mathscr{H}_{\rm QS}|i\right>+\sum_{j \neq i}\frac{|\left<j|\mathscr{H}_{\rm QS}|i\right>|^2}{E_j^{(0)}-E_i^{(0)}}\nonumber\\
&=&E_i^0+g_\Gamma \epsilon_{\Gamma} \left<i| O_{\Gamma} |i\right>+g_\Gamma^2 \epsilon_{\Gamma}^2 \sum_{j \neq i}\frac{|\left<j| O_{\Gamma} |i\right>|^2}{E_j^{(0)}-E_i^{(0)}}
\end{eqnarray}

where $E_i$ is the energy in the quantum state $\left| i \right>$. The total free energy $F$ in a crystal is the sum of the elastic energy and the free energy of the electronic system,

\begin{eqnarray}
F = \frac{1}{2} \sum_{\Gamma} C_{\Gamma}^{(0)}\epsilon_{\Gamma}^2-Nk_{\rm B}T\ln \sum_{i}\exp\{-E_i(\epsilon_\Gamma)/k_{\rm B}T\},
\end{eqnarray}

where $C_{\Gamma}^{(0)}$ is the elastic constant without quadrupole-strain interaction, and $N$ is the number of electric quadrupoles, {\it i.e}., single atomic vacancies, per unit volume in the case of Si. The elastic constant $C_{\Gamma}(T)$ is obtained by the second derivative of the Helmholtz free energy $F$ with respect to strain in the limit $\epsilon_\Gamma \rightarrow 0$.
\begin{equation}
C_\Gamma(T)=\biggl( \frac{\partial^2 F}{\partial \epsilon_{\Gamma}^2}  \biggr)_{\epsilon_\Gamma \rightarrow 0} = C_\Gamma^0-Ng_\Gamma^2\chi_\Gamma(T),
\end{equation}
\begin{eqnarray}
-g_\Gamma^2\chi_\Gamma&= \left<\frac{\partial^2E_i}{\partial\epsilon_\Gamma^2}\right>
-\frac{1}{k_{\rm B}T}\Bigl[\left<\Bigl(\frac{\partial E_i}{\partial\epsilon_\Gamma}\Bigr)^2 \right>-\left< \frac{\partial E_i}{\partial\epsilon_\Gamma}\right>^2\Bigr],
\end{eqnarray}
where $\chi_\Gamma (T)$  is the quadrupolar susceptibility of $\Gamma$ symmetry.
In addition to the strain-quadrupole interaction, inter-site quadrupole-quadrupole interactions are considered. Here, we focus on the $k$ site. Quadrupoles other than the $k$ site are taken into account by mean-field approximation,

\begin{equation}
\mathscr{H}_{\rm QQ (\Gamma)}^{MF}=-\sum_{k}g'_{\Gamma}\left<O_{\Gamma}\right>O_{\Gamma}^{(k)},
\end{equation}

and the Hamiltonian of the system can be written as:

\begin{eqnarray}
\mathscr{H}&=&\mathscr{H}_{0}+\mathscr{H}_{QS}+\mathscr{H}_{QQ}^{MF}\nonumber\\
&=&-g_{\Gamma}\sum_{k}\sum_{\Gamma}O_{\Gamma}^{(k)}\epsilon_{\Gamma}^{eff}.
\end{eqnarray}

$\epsilon_{\Gamma}^{eff}=\epsilon_{\Gamma}+\frac{g'_{\Gamma}}{g_{\Gamma}} \left<O_{\Gamma}\right> $is the effective mean-field strain. The elastic constants [Eq. (18)], including inter-site interactions, are rewritten as:

\begin{equation}
C_\Gamma(T)=C_\Gamma^0-\frac{Ng_\Gamma^2\chi_\Gamma(T)}{1-g'_\Gamma\chi_\Gamma(T)}.
\end{equation}

The second-order perturbation term in the energy [the third term on the right-hand side of Eq. (16)], and the so-called van Vleck term in the quadrupolar susceptibility [the first term on the right-hand side of Eq. (19)] should also be addressed. In first-order perturbation, the thermal average due to strain is zero. So, only the Curie term, which is inversely proportional to $T$, appears in the quadrupolar susceptibility,

\begin{eqnarray}
-g_\Gamma^2\chi_\Gamma &=&-\frac{1}{k_{\rm B}T}\Bigl[\left<\Bigl(\frac{\partial E_i}{\partial\epsilon_\Gamma}\Bigr)^2 \right>-\left< \frac{\partial E_i}{\partial\epsilon_\Gamma}\right>^2\Bigr].\nonumber\\
\end{eqnarray}

Diagonalizing the quadrupole operators with $\Gamma$ symmetry in Eq. (13) and calculating the quadrupolar susceptibility with $\Gamma$ symmetry in Eqs. (14) to (22), we obtain,

\begin{equation}
-g_\Gamma^2\chi_\Gamma =-\frac{2g_\Gamma^2}{3k_{B}T}.
\end{equation}

Substituting Eq. (24) into Eq. (22) yields

\begin{equation}
C_\Gamma(T)=C_\Gamma^{(0)}\Bigl( \frac{T-T_C}{T-\Theta} \Bigr).
\end{equation}

Here, $\Theta = \frac{2g'_\Gamma}{3k_{B}}$ is the coupling constant of the quadrupole-quadrupole interaction. $T_C$ is the total energy of the quadrupolar interaction, which is described as $T_C = \Theta + \Delta_{JT}$, with the binding energy of the strain field between quadrupoles $\Delta_{JT} = T_C - \Theta = \frac{2Ng_\Gamma^2}{3k_{B} C_{\Gamma}^{(0)}}$, $i.e.$, the so-called Jahn-Teller energy. Using the transformational binding energy $\delta_{\Gamma}= \sqrt{\frac{2}{3}} g_{\Gamma}$, the quadrupole susceptibility can be rewritten as

\begin{equation}
\Delta_{JT} = \frac{N \delta_{\Gamma}^2}{C_{\Gamma}^{(0)}}.
\end{equation}

Hence, the concentration $N$ of defect sites possessing electric quadrupole moments is obtained as follows:

\begin{equation}
N = \frac{\Delta_{JT} C_{\Gamma}^{(0)}}{\delta_{\Gamma}}.
\end{equation}

From Eq. (16), the quadrupole-strain coupling constant $g_\Gamma$ and the intersite quadrupole coupling constant $g'_\Gamma$ have the following relationship with the effective charge $Z^*$ and the effective radius of the quadrupole $a^*$:

\begin{eqnarray}
g_\Gamma &=& k_{\Gamma} Z^* a^{*2}\nonumber\\
g'_\Gamma &=& k'_{\Gamma} Z^{*2} a^{*4}.
\end{eqnarray}

where $k_{\Gamma}$ and $k'_{\Gamma}$ are proportionality constants. From the above discussion, we obtain the following relationship between the deformation coupling energy $\delta_{\Gamma}, Z^*$, and $a^*$:

\begin{eqnarray}
\delta_{\Gamma}= \sqrt{\frac{2}{3}} g_{\Gamma} = \sqrt{\frac{2}{3}} k_{\Gamma} Z^*a^*2
\end{eqnarray}

\newpage
\vspace{10mm}
  \section*{\label{sec:Schematic}5. COMPARISON OF Si AND DIAMOND }
The energy gap between the valence and conduction bands for diamond is 5.45 eV, which is approximately five times larger than the 1.17 eV for Si with diamond structure as well. Thus, diamond is expected to be the ultimate semiconductor for outstanding high-power and high-frequency performance~\cite{Kasu2021}. Other major properties are compared in the dyad and shown in Table AII.

\begin{table*}[htbp]
\centering{
\caption{Comparison of various parameters between Si and diamond~\cite{Rikanenpyou}.}
\vspace{5mm}
\begin{tabular}{|l|c|c|}
\hline
~~&Si&Diamond\\
\hline
Lattice constant: $a$~&5.43 \AA&3.567 \AA\\
\hline
Nearest-neighbor distance: $\frac{\sqrt{3}a}{4}$~&2.35 \AA&1.55 \AA\\
\hline
Melting point~&1415 $^\circ$C (1 atm)&$\sim$3550 $^\circ$C (1 atm)\\
\hline
Density: $\rho$~&2.33 g/cm$^3$&3.515 g/cm$^3$\\
\hline
Band gap~&1.12 eV&5.45 eV\\
\hline
Cohesive energy (eV per atom)~&4.63\footnote{See, M. Schl\"uter, Proc. Int. School of Physics ``Enrico Fermi'', p.495 (1985),  and references therein.}&7.37\\
\hline
Deformation binding energy: $\delta_{\Gamma}^0$~&$2.28 \times 10^5$ K\footnote{coupling constant of Boron-doped Si ~\cite{Okabe2013}.}& ? \\
\hline
Magnitude of low-temperature softening~&$1.6 \times 10^{-4}$ (160 ppm)\footnote{non-doped CZ-grown Si~\cite{Goto2007}.}&$1.25 \times 10^{-4}$ (125 ppm)
\footnote{Type IIa HPHT diamond (present work).}\\
(relative change in $\Delta C_{44}/C_{44})$~&~&~\\
\hline
Jahn-Teller energy: $\Delta_{JT}$~&0.51 mK&0.046 mK\\
\hline
\end{tabular}
}
\vspace{5mm}
\end{table*}

\newpage
  \section*{\label{sec:Schematic}6.	MAGNETIC-FIELD DEPENDENCE OF THE SOFTENING OF $C_{44}$ }
The temperature dependence of the elastic constant $C_{44}$ in the three types of diamonds, as shown in Fig. 3 of the main text, was studied as well in magnetic fields, as presented in Fig. A2. The results reveal that the HPHT type-Ib and type-IIa diamonds show small differences of $C_{44}$ at 0 T and in a magnetic field of 16.5 T. However, CVD type-IIa diamond showed a slightly changed temperature dependence below 200 mK in a magnetic field of 14 T. This difference should be addressed in future investigations.

\begin{figure}[h]
\includegraphics[width=0.7\linewidth]{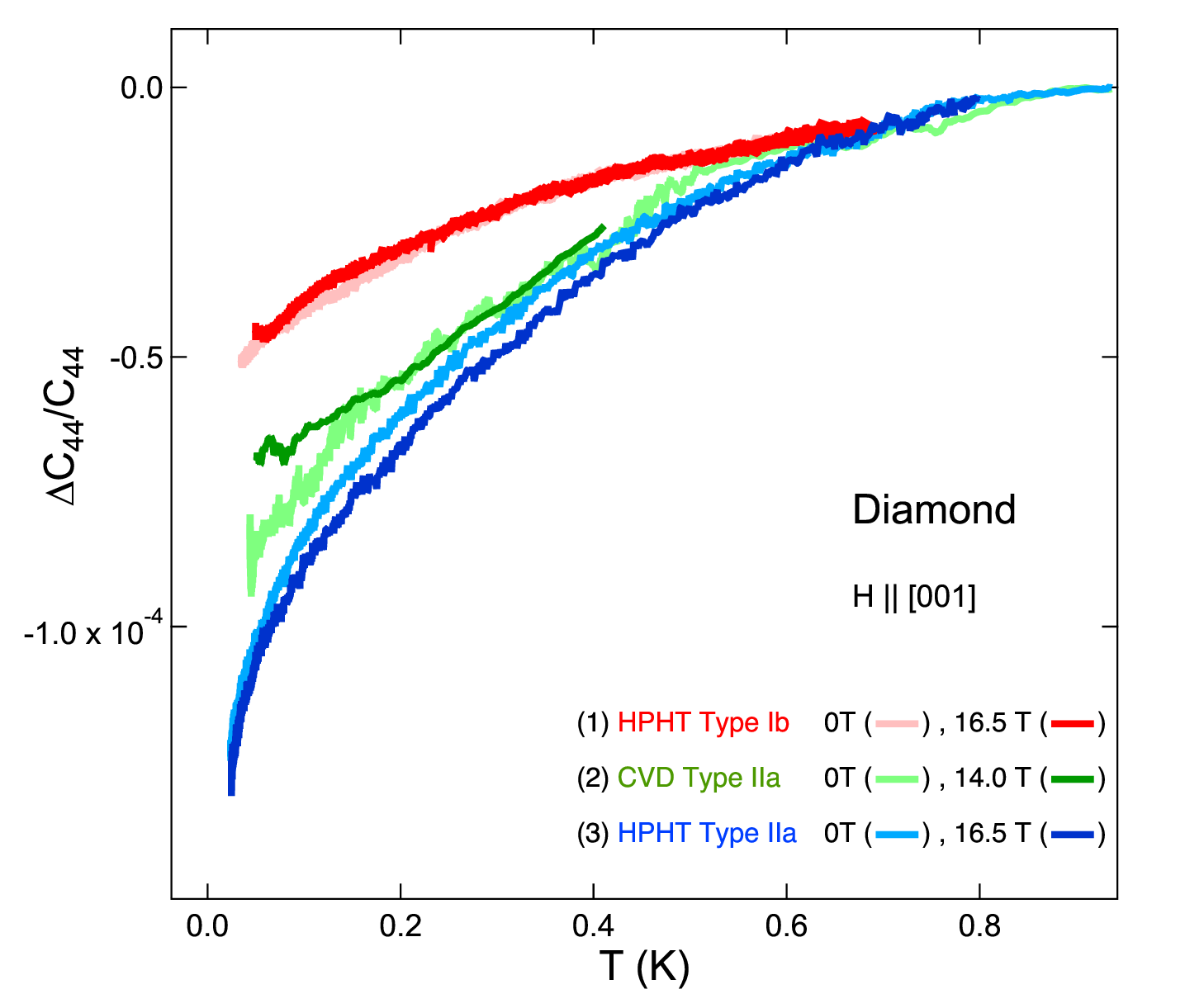}\\
{
FIG. A2~~Elastic softening of $C_{44}$ for three different types of diamond at zero and magnetic fields of 16.5 or 14 T, applied for $H \|$ [001].}
\end{figure}

 \section*{\label{sec:Schematic}7. EFFECTS OF NV$^-$, NV$_0$,  AND NUCLEAR SPINS }
 
In an NV center, the vacancy site exhibits reduced $C_{3v}$ symmetry due to the substitution of one C atom by N, leading to a local symmetry breaking. Therefore, the $sp^3$ orbital splits into two $a_1$ singlets and one $e$ doublet~\cite{Cox1994, Doherty2013}. When the nitrogen donor in the substituted position provides the carrier for the valence 1, the NV center is negatively charged (NV$^-$). According to Hund’s first rule, the ground state adopts the highest value of total spin $S$, which is consistent with the Pauli exclusion principle. The six electrons occupy the $a_1$ singlets and the $x$ and $y$ orbits of the $e$ doublet to satisfy the condition that all wavefunctions are antisymmetric. Then, due to the spin-orbit interaction, the ground state of NV$^-$ results in $S = 1$ or $S = 0$ ground state with irreducible representation $^3$A$_2$, or $^1$A$_1$. No electric quadrupole degrees of freedom exist in the $^1$A$_1$ singlet. Therefore, the $e$ orbitals of NV$^-$ do not contribute to the softening of $C_{44}$. 

In this context, we also examine further possible scenarios. Stable isotopes of C include $^{12}$C, $^{13}$C, and $^{14}$C, with natural abundances of 98.9\%, 1.1\%, and 1.2 $\times 10^{-8}$\% and nuclear spin quantum numbers $I = 0, 1/2$, and 0, respectively. None of the isotopes has a nuclear quadrupole moment; hence, the elastic constants remain unaffected at zero magnetic field. The elastic constant depends hardly on magnetic fields and the change is negligibly small compared to the low-temperature softening. Thus, we conclude that the effect of $^{13}$C is small and that nuclear spins need not to be considered. 

\end{document}